\documentclass[12pt]{article}%
\usepackage{amsmath}
\usepackage{graphicx}%
\usepackage{amsfonts}%
\usepackage{amssymb}
\newtheorem{theorem}{Theorem}
\newtheorem{acknowledgement}[theorem]{Acknowledgement}

\begin{document}

\author{C. S. Unnikrishnan\thanks{E-mail address: unni@tifr.res.in}\\\textit{Gravitation Group, Tata Institute of Fundamental Research,}\\\textit{Homi Bhabha Road, Mumbai - 400 005, India \&}\\\textit{NAPP Group, Indian Institute of Astrophysics,}\\\textit{Koramangala, Bangalore - }560 034, India}
\title{Spin-Statistics connection and the gravity of the Universe: The Cosmic
connection }
\date{}
\maketitle

\begin{abstract}
I suggest that the Spin-Statistics connection is a consequence of the phase
shifts on quantum scattering amplitudes due to the induced gravitomagnetic
field of the whole Universe at critical density. This connection was recently
brought out in the context of a new theory of relativity in flat space with
matter, called Cosmic Relativity. This prediction of the correct gravitational
phases is a consequence of any relativistic gravitation theory in the presence
of the massive Universe. This can also be interpreted as related to the Mach's
principle applied to quantum phenomena. Perhaps this is the simplest valid
proof of the Spin-Statistics Theorem, and it finally identifies the physical
origin of the connection.

\end{abstract}

\section{General remarks}

The main result of this paper is the suggestion that the well known
spin-statistics connection is a consequence of the gravitational interaction
of the quantum system with the entire Universe. In other words, the Pauli
exclusion is a consequence of the relativistic gravitational interaction with
the critical Universe, which is always present. Motion relative to masses in
the Universe generates gravitational vector potentials that interact with the
spin. The curl of this potential is a magnetic analog in gravity, and the
spin-gravitomagnetic interaction energy leads to phase changes in the quantum
state. By considering the difference in phases induced by the gravitomagnetic
interaction in two possible amplitudes for scattering of identical particles
we are able to deduce the spin-statistics connection; identical integer spin
particles obey Bose-Einstein statistics and identical half-integer spin
particles obey Fermi-Dirac statistics.

This result might answer for the first time the long-standing query -- what
are the physical reasons behind the spin-statistics connection? It also
answers why the connection is valid in non-relativistic, two-particle
situations despite the general impression that it is a consequence of
relativistic field theory.

The fundamental conceptual basis that underlies the physical effects discussed
here is the theory of Cosmic Relativity, discussed in detail in gr-qc/0406023
\cite{cosrelx}. There is also a discussion of the cosmic connection between
gravity and spin dynamics in that paper. Here we focus on the spin-statistics
connection exclusively, discuss the physical and mathematical arguments that
links this to the gravity of the Universe, and also discuss briefly some
aspects in the context of other proofs of the spin-statistics connection.

In Cosmic Relativity, a particle that is moving in the cosmic gravitational
potential $\phi$ of the Universe experiences a modified gravitational
potential and a vector gravitational potential equal to
\begin{align}
\phi^{\prime}  &  =\phi(1-V^{2}/c^{2})^{-1/2}\\
A_{i}  &  =\phi\frac{V_{i}}{c}(1-V^{2}/c^{2})^{-1/2}%
\end{align}
For Universe with critical density, the quantity $\phi=1$ (in unit of $c^{2}%
$). In the more rigorous and exact language of the metric for the massive
Universe, this is equivalent to writing $g_{0i}=\frac{V}{c}(1-V^{2}%
/c^{2})^{-1/2}.$ Circular motion then gives a nonzero curl for the velocity
field, and this is a gravitomagnetic field due to the entire Universe,%

\begin{equation}
\overrightarrow{B}_{g}=\nabla\times\mathbf{A}=\Omega(1-v^{2}/c^{2})^{-1/2}%
\end{equation}
This gravitomagnetic field couples to the spin angular momentum. The modified
potential modifies precession rate of spin, and also modifies quantum phases
of states with spin. Though I have written the gravitomagnetic field in terms
of the post-Newtonian potentials, to clarify the physical issues in terms of
familiar concepts like an interaction energy of the quantum spin in a
gravitomagnetic field, the final expressions concerning `phase' of quantum
amplitudes are expected to be correct to all orders. This is because the phase
is the product of the interaction energy and time duration of interaction, and
these two variables are affected by gravity of the Universe in an exactly
reciprocal way.

I have been convinced of a connection between quantum spin dynamics in the
Universe and effects like Thomas precession, Spin-Statistics connection, and
geometric phases of fundamental particles taken in closed trajectories in
momentum space since about 1995. Around the same time, it also became clear to
me that a consistent formulation of relativistic effects in a Universe with an
operational way of precisely determining the `absolute velocity' relative to a
preferred rest frame (the CMBR), and also the possibility of an absolute time
(temperature of the CMBR)\ can be done only by taking the gravity of the
Universe as the major physical reason for all those effects usually attributed
to kinematics \cite{unni-twin}. But, it was a major psychological barrier to
think that special relativity based on mere kinematics in empty, matter-free
space then will have to be \emph{replaced} by a new theory of relativity based
only on gravitational effects of the matter-filled Universe. The two ideas go
together, inseparably. Finally, such a new theory is proposed and discussed in
detail in ref. \cite{cosrelx}. I have previously discussed Thomas precession
as an effect of spin-gravitomagnetic field coupling, and the spectral fine
structure splitting as due to the gravitational interaction of the quantum
spin with the Universe \cite{unni-mpla,unni-erice}.

\section{The Spin-Statistics Connection}

The spin-statistics connection in quantum mechanics is a very intriguing fact.
It is one of the most basic invariant facts of the physical world, yet we
understand it only in terms of certain consistency relations between
mathematical constructs and not in terms of a deeper physical aspect related
to properties of the quantum spin angular momentum and interaction between
particles. This is evident in a quotation from Feynman \cite{fey}, ``Why is it
that particles with half-integer spin are Fermi particles whose amplitudes add
with the minus sign, whereas particles with integral spin are Bose particles
whose amplitudes add with the positive sign? We apologize for the fact that we
cannot give you an elementary explanation. ..has been worked out by Pauli from
complicated arguments of quantum field theory and relativity\ldots It appears
to be one of the few places in physics where there is a rule that can be
stated very simply, but for which no one has found a simple and easy
explanation\ldots This probably means that we do not have a complete
understanding of the fundamental principle involved.''

While there is no physical understanding of the connection, mathematical
proofs invoking field theoretic reasoning exist. The proof by Pauli used the
mathematical fact that the quantization of the field of integer spin particles
is associated with commutator relations and the quantization of half-integer
spin particles is associated with anti-commutator relations between operators,
and these coupled with requirements from Lorentz invariance provided the proof
\cite{pauli}. Many rigorous proofs with assumptions on properties of quantum
fields, and Lorentz invariance exist, and a proof by Schwinger relied on
requirement of invariance under time reversal. See ref. \cite{sud} for details
with very instructive and readable commentary. There have been many attempts
to provide simple proofs, and the general assessment seems to be that there is
no simple proof, let alone a physical understanding of the connection. There
have been some discussion on whether there is any `simple proof' and whether
the claimed simple proofs are physically defendable, in the American Journal
of Physics, initiated by a question by D. E. Neuenschwander \cite{neuen}. This
discussion motivated a thorough discussion by I. Duck and E. C. G. Sudarshan
on many aspects of the simple proofs of the connection, and also a valuable
discussion of many aspects of both the `hard proofs' and the simple proofs in
their recent book, \textit{Pauli and the Spin Statistics Theorem} \cite{sud}.

Before proceeding further I state the spin-statistics connection:

a) Particles with integer spin are bosons and they obey the Bose-Einstein statistics.

b) Particles with half-integer spin are fermions and they obey the Fermi-Dirac statistics.

Equivalently,

a) The amplitude for a scattering event between identical integer spin
particles and the amplitude with an exchange of the particles add with a plus
($+$) sign. In other words, the phase difference between the scattering
amplitude and the exchanged amplitude is an integer multiple of $2\pi.$

b) The amplitude for a scattering event between identical half-integer spin
particles and the amplitude with an exchange of the particles add with a minus
($-$) sign. In other words, the phase difference between the scattering
amplitude and the exchanged amplitude is an odd integer multiple of $\pi.$

A geometric understanding of these statements was published by Berry and
Robbins \cite{berry}, and several authors have invoked the relation between
rotation operators and exchange of particles in quantum mechanics to prove the
spin-statistics theorem \cite{spinstat-sud}. Sudarshan has been arguing for
the existence of a simple proof that is free of arguments specific to
relativistic quantum field theory\cite{sud}. While these attempts have
clarified several issues regarding the connection, none provides a good
physical understanding of the connection.

It may be noted that physically the connection is applicable for any two
identical particles, in non-relativistic quantum mechanics. Thus we should
expect that the physical proof need not depend on relativistic quantum field
theory. In any case, there are certain minimal expectations for a physically
valid proof of the spin-statistics connection: 1) It should of course derive
the correct phase factors between the two possible scattering amplitudes,
direct and exchanged, from a physical interaction, or from a geometro-physical
reasoning affecting only the two particles, 2) the phase difference should be
independent of the specific interaction between the two particles, 3) if there
is any modification to the exactness of the phase difference (zero for bosons
and $\pi$ for fermions), then it should be small enough not to affect observed
facts like stability of atoms, black-body spectrum etc.

Consider the scattering of two identical particles, Fig. 1.%

\begin{figure}
[ptb]
\begin{center}
\includegraphics[
height=2.2355in,
width=4.6432in
]%
{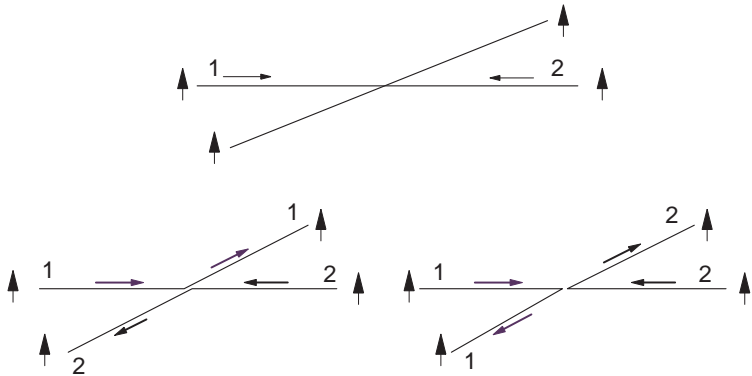}%
\caption{Amplitudes for scattering between indistinguishable particles}%
\end{center}
\end{figure}

The upper event can happen by two quantum amplitudes, shown in the lower
panel. The amplitudes for these two processes differ by only a phase for
identical particles in identical states, since the two processes are
indistinguishable. The particles are assumed to be spin polarized in identical
directions, perpendicular the plane containing the scattering event. Only
then, the initial and final states and the two processes are
indistinguishable. We note the important fact that \emph{the two amplitudes
have to be considered always in the gravitational field of the entire
Universe}. The task is to calculate the phase difference between the two
amplitudes due the gravitational interaction with the entire Universe. We can
calculate the phase changes in any configuration, but at present we want to
discuss only the phase difference for indistinguishable states.

First we note that the two processes are different in the angle through which
the \emph{momentum vector} of the particles turn in the scattering process. In
fact that is the only difference between the two amplitudes. The difference in
angles is just $\pi.$ (This is why it is equivalent to an exchange - what is
really exchanged is the momentum vector after the scattering). The calculation
for each particle can be done by noting that the $k$-vector can be considered
without deflection, but the entire Universe turned through an appropriate
angle, with angular velocity of this turning decided by the rate of turning of
the $k$-vector. It does not matter whether we are dealing with massless
particles or massive particles since the only physical fact used is the change
in the direction of the $k$-vector of the particle.

Figure 2 depicts the `motion of the Universe' as seen from the rotating frame
at the point where the scattering takes place. \ The part `A' corresponds to
the mass currents relative to each of the scattering particle when the
scattering angle is as shown in the upper diagram. The other possible
amplitude then happens with the scattering angle that is larger by $\pi,$ and
the mass currents act for a larger duration, as in `B'. In any relativistic
theory of gravity such mass currents generate a Lense-Thirring field, or
equivalently a gravito-magnetic field. This is equal to the curl of the vector
potential generated by the mass currents, in the post-Newtonian language, and
equivalently the `curl' of the metric coefficient for the term $dx~dt$ in the
more rigorous metrical description.%

\begin{figure}
[ptb]
\begin{center}
\includegraphics[
height=1.9951in,
width=3.5492in
]%
{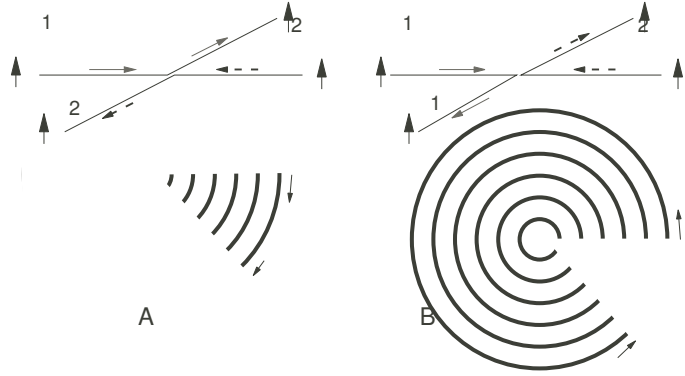}%
\caption{The curved paths of distant galaxies relative to the turning momentum
vector at the quantum scattering event. The large mass currents generate large
gravitomagnetic field which couples to spin $s$ and affects the energy and
phase of the quantum state. A and B show the real physical difference of the
two amplitudes relative to the massive Universe. The change in phase in a
critical Universe is simply $s\theta,$ where $\theta$ is the angle through
which the momentum vector of each particle turns during scattering.}%
\end{center}
\end{figure}

The phase change in the state of each particle is given by the product of the
duration of interaction and the spin-gravitomagnetic interaction energy,
\begin{equation}
\varphi=\mathbf{s}\cdot\overrightarrow{\Omega}(1-V^{2}/c^{2})^{-1/2}\times
t(1-V^{2}/c^{2})^{1/2}=s\theta
\end{equation}
where $\theta$ is the angle through which the $k$-vector has turned in the
scattering process. Note that this is a gravitational phase shift due to the
relativistic gravitational interaction of the quantum spin with the entire
causally connected Universe, and it involves all the fundamental constants
$G,$ $c$ and $\hbar.$ These constants are hidden from the final expression
because $\sum_{i}Gm_{i}/c^{2}r_{i}=1$ in a critical Universe. When the density
is not critical, these constants will appear explicitly in the expression for
this phase change. Also note that this phase difference is an unambiguous
prediction from any theory of relativistic gravity in a Universe with critical density.

The momentum vectors turn in the same sense for both the particles and
therefore the total phase change is $\varphi_{1}=2s\theta_{1}$ where
$\theta_{1}$ is the angle through which the $k$-vector turns for the first
amplitude. For the second amplitude the phase change is
\begin{equation}
\varphi_{2}=2s\theta_{2}%
\end{equation}
The phase difference between the two amplitudes is
\begin{equation}
\Delta\varphi=2s(\theta_{1}-\theta_{2})=2s\times\pi
\end{equation}

Now we can see how the spin-statistics connection is related to the phase
changes due to the gravitational interaction of the spin with the Universe.
For zero-spin particles the proof is trivial since there is no spin coupling
to the Universe,
\begin{equation}
\Delta\varphi=s(\theta_{1}-\theta_{2})=0\times2\pi=0
\end{equation}
and therefore zero-spin particles are bosons and their scattering amplitudes
add with a $+$ sign.

It is sufficient to show the connection for spin-$1/2$ particles and the
higher spin cases can be constructed from spin-half using the Schwinger
construction \cite{sud}. The only case to be treated when dealing with
identical, indistinguishable states of spin is the one in which the spins are
identically pointing, perpendicular to the plane containing the $k$-vectors.
The phase angle difference between the two amplitudes then is%
\begin{equation}
\Delta\varphi=2s(\theta_{1}-\theta_{2})=2\times\frac{1}{2}\times\pi=\pi
\end{equation}
The relative phase is $\exp(i\pi)=-1.$ The amplitudes add with a negative
sign. Therefore, half-integer spin particles obey the Pauli exclusion and the
Fermi-Dirac statistics.

Two features of this proof may be noted immediately; 1) scattering of
\emph{any} two particles in a multiparticle system of identical fermions
introduces a minus ($-$) sign between the original amplitude and the exchanged
amplitude for the total system, 2) the proof is valid for interacting
particles since the phase changes due to interactions are identical for the
two amplitudes, as all other dynamical phases in the relevant diagrams. The
only phase difference is due to the rotation of the momentum vector in the
presence of the Universe.

\section{Concluding remarks}

The remarkable connection between quantum physics and gravity of the Universe
is indeed startling. But this cosmic connection is also a natural consequence
in a critical Universe in which everything is gravitationally interacting with
everything else. It is satisfying to see that a deep physical phenomenon is
linked to a universal physical interaction and not just to mathematical
structures and consistency.

Many of the measured geometric phases on particles with spin, when they are
taken around trajectories in space with their momentum vector turning in
space, are of the same nature. Those geometric phases are simply given by
\begin{equation}
\varphi=\oint sd\theta
\end{equation}
where I mean an integral over the entire trajectory with appropriate signs.
This may suggest the relation between our proof and the discussion by Berry
and Robbins \cite{berry}. It also clarifies why the proofs that depend on
rotational properties of wavefunctions have some success.

By identifying the physical origin as the gravitational interaction, we can
now calculate the phase change for the two amplitudes for any \emph{arbitrary
initial and final states}, and thus we have the most general statement about
the relative phase between the amplitude of any such scattering event. It
might be possible to extend this physical insight further and arrive at the
spin-statistics connection without specifically referring to scattering amplitudes.

In discussions of the proofs of the spin-statistics connection it is argued
that the phase factor is either $+1$ or $-1,$ by requiring that the system
comes back to the same state, including phase, after two exchanges. One
exchange of the particles converts the two amplitudes into each other. However
such assumptions have been criticized with good reasons \cite{spinstat-sud}.
We do not use such an assumption for establishing the spin-statistics
connection from the gravitational interaction, and this is a great advantage.
In fact, the correct phase factors emerge naturally for each particle
independently as a dynamical phase that depends on the interaction of the spin
of each particle with the Universe. The calculation is applicable to any spin,
and does not by itself rule out other fractional spin particles and
statistics, unless an additional assumption is made about the complete
equivalence, including phase, of two-particle configuration after two
exchanges. So, in our case, the standard spin-statistics connection can be
shown without this assumption, and with this assumption it seems possible to
show that there can only be half-integer and integer spin particles in this
Universe. Duck and Sudarshan note that all proofs of the spin-statistics
connection are `negative proofs' showing the impossibility of the ``wrong''
statistics. The proof relating the gravity of the Universe and the statistics
is a `positive proof' explicitly calculating the relevant phase differences.

Finally, there is no doubt that the cosmic connection discussed here is
physically simple and transparent, and the resulting proof is simple. The
issue is only whether this exclusively is the cause of the spin-statistics
connection. If it is, it certainly answers satisfactorily Neuenschwander's
\cite{neuen} query I referred to earlier. Whether or not this is the only
physical aspect involved in the spin-statistics connection the phase changes
we discussed are unavoidable consequence of quantum dynamics in this massive
critical Universe. Therefore, they surely have a role to play in the phase
differences of the scattering amplitudes we discussed, and hence in the
spin-statistics connection.

\bigskip

\begin{acknowledgement}
I thank A. P. Balachandran and E. C. G. Sudarshan for valuable discussions. I
thank Anil Shaji for clarifying the criticisms on the so called simple proofs.
\end{acknowledgement}

\end{document}